\documentclass[aps,prl,twocolumn]{revtex4}
\usepackage{amsmath}

\begin{document}
\title{Comment on ``Explicit Analytical Solution for Random Close
  Packing in $d=2$ and $d=3$''}

\author{W. Till Kranz}
\affiliation{Institut f\"{u}r Theoretische Physik, Universit\"at zu K\"oln, 50937 K\"{o}ln, Germany}

\date{\today}

\maketitle

In a recent letter \cite{zaccone22} \citeauthor{zaccone22} offered a
fresh look at the jamming transition which garnered a number of
comments challenging the plausibility of its predictions
\cite{duyu+ni22,charbonneau+morse22,blumenfeld22}. Unfortunately, the
article already contains two mathematical flaws, not addressed in the
erratum \cite{zaccone22e}, that render the proposed approach to
theoretically determine the jamming density $\phi_{\mathrm{RCP}}$
unfeasible.

As the contact value $g(\sigma)$ of the pair correlation function
$g(r)$ diverges at close packing \cite{charbonneau+kurchan17}, Eq.~(6)
inserted in Eq.~(5) in Ref.~\cite{zaccone22} appear suspicious,
\begin{equation}
  \label{eq:1}
  g(r) = g_0g(\sigma)\delta(r - \sigma) + g_{\mathrm{BC}}(r),
\end{equation}
especially if contrasted with the established expression
\cite[Eq.~(17)]{torquato18}
\begin{equation}
  \label{eq:2}
  g(r) = \frac{z}{\rho_{\mathrm{CP}}\mu(\sigma)}\delta(r - \sigma) 
  + g_{\mathrm{BC}}(r).
\end{equation}
Here $\mu(\sigma)$ is the surface area of a $d$-dimensional sphere of
radius $\sigma$, $z$ is the coordination number, and
$\rho_{\mathrm{CP}}$ is the close packing density. Note that the
prefactor of the delta distribution is finite in Eq.~(\ref{eq:2}) and
infinite in Eq.~(\ref{eq:1}). 

To clarify this issue, let's start at a packing fraction
$\phi \equiv \rho v(\sigma/2) < \phi_{\mathrm{CP}}$, lower than the
close packing value $\phi_{\mathrm{CP}}$. Here $v(\sigma/2)$ denotes
the volume of a $d$-dimensional sphere of diameter $\sigma$. In
anticipation of the equation in question and the limit
$\phi\to\phi_{\mathrm{CP}}$, let us split $g(r)$ into a divergent part
and a bounded regular part
$g(r) = g_{\mathrm{div}}(r/\sigma) + g_{\mathrm{BC}}(r/\sigma)$
\footnote{Note, that I had originally termed the divergent
  contribution \emph{singular}. As has been highlighted by Zaccone
  \cite{zaccone22r}, the notion of a \emph{singular probability
    density} has a different definition not fulfilled and not needed
  here. Instead, the crucial property of $g_{\mathrm{div}}$ is its
  \emph{divergence} with $g(\sigma)$.}. Let us choose the splitting
such that $g_{\mathrm{BC}}(r\to\sigma) \to 0$ and
$g_{\mathrm{div}}(r/\sigma) = g_0g(\sigma)\tilde G(r/\sigma -
1)$. Here
\begin{equation}
  \label{eq:3}
  g_0 := \int_1^{\infty}dxg_{\mathrm{div}}(x)/g(\sigma) 
\end{equation}
is chosen such that $\int_0^{\infty}dx\tilde G(x) = 1$. Note that
$g_{\mathrm{div}}(r\searrow\sigma) \equiv g(\sigma)$. At variance
with Ref.~\cite{zaccone22}, $g_0 = g_0(\phi)$ will in general be a
function of the packing fraction $\phi$. Rescaling the argument,
$\tilde G(x) \equiv G(xg(\sigma))$, the term $g(\sigma)G(xg(\sigma))$
defines a delta series as $g(\sigma)\to\infty$ \cite{kanwal98}, such
that the contact contribution of Ref.~\cite{zaccone22},
$g_c(r) \equiv
\lim_{\phi\nearrow\phi_{\mathrm{CP}}}g_{\mathrm{div}}(r/\sigma)$,
can be calculated. We find
\begin{align}
  g_c(r) &= \lim_{g(\sigma)\to\infty}g_0g(\sigma)
           G\left((r/\sigma - 1)g(\sigma)\right)\\  \label{eq:4}
         &= g_0^{\mathrm{CP}}\delta(r/\sigma - 1) = g_0^{\mathrm{CP}}\sigma\delta(r -
           \sigma),
\end{align}
where $g_0^{\mathrm{CP}} \equiv g_0(\phi\to\phi_{\mathrm{CP}})$. This shows
that the spurious divergent factor $g(\sigma)$ in Eq.~(6) of
Ref.~\cite{zaccone22} has to be absorbed into the delta function.

Using Eq.~(\ref{eq:4}) in Eq.~(\ref{eq:2}), a general relation between
$g_0^{\mathrm{CP}}$ and the coordination number $z$ can be established
in any dimension $d\geq2$,
$z = \rho_{\mathrm{CP}}\mu(\sigma)g_0^{\mathrm{CP}}\sigma$
\footnote{Note, however, that Zaccone \cite{zaccone22r}, questions the
  compatibility of Eq.~(\ref{eq:4}) with Eq.~(\ref{eq:2}).}. Using
$\mu(\sigma)\sigma/v(\sigma/2) = d2^d$, we can express the relation
more succinctly as $z = d2^d\phi_{\mathrm{CP}}g_0^{\mathrm{CP}}$.

Imposing the isostaticity condition $z_{\mathrm{iso}} = 2d$, the
normalization constant,
\begin{equation}
  \label{eq:6}
  g_0^{\mathrm{iso}} = \frac{1}{2^{d-1}\phi_{\mathrm{RCP}}},
\end{equation}
is not universal but directly related to the random close packing
density $\phi_{\mathrm{RCP}}$. Following Ref.~\cite{zaccone22} and
assuming $\phi_{\mathrm{RCP}} = 0.60$ -- $0.69$ in $d=3$, we have
$g_0^{\mathrm{iso}}(d=3) = 0.36$ -- $0.42$. We may also
determine $g_0$ for the fcc-crystal in $d=3$ with $z=12$ and the
crystal close packing fraction $\phi_{\mathrm{fcc}} = \pi/\sqrt{18}$,
\begin{equation}
  \label{eq:7}
  g_0^{\mathrm{fcc}} = \frac{1}{2\phi_{\mathrm{fcc}}} =
  \frac{\sqrt{18}}{2\pi}
  = 0.675\ldots
\end{equation}
The protocol and density dependence of $g_0$ clearly precludes the
calibration $g_0^{\mathrm{iso}}(d=3) \equiv g_0^{\mathrm{fcc}}$
assumed in Ref.~\cite{zaccone22}.

Equations (\ref{eq:3}) and (\ref{eq:6}) yield the interesting
constraint,
\begin{equation}
  \label{eq:8}
  2^{d-1}\phi_{\mathrm{RCP}}
  \int_1^{\infty}dxg_{\mathrm{div}}^{\mathrm{iso}}(x)/g^{\mathrm{iso}}(\sigma) = 1,
\end{equation}
that relates the random close packing density to an integral over the
normalized divergent part of the pair correlation function. Analyzing
this constraint is beyond the scope of this comment. However, it is
expressly independent of the pair correlation's contact value
$g(\sigma)$. Besides the non-universality of $g_0$'s value, this
spoils the approach put forward in Ref.~\cite{zaccone22}, whose beauty
lies in relating $\phi_{\mathrm{RCP}}$ to $g(\sigma)$, or,
equivalently, the equation of state.

To conclude, Eq.~(\ref{eq:4}) in agreement with Eq.~(17) in
Ref.~\cite{torquato18} does not contain $g(\sigma)$ and corrects
Eq.~(6) in Ref.~\cite{zaccone22}, where the explicit dependence on
$g(\sigma)$ is crucial for the subsequent derivations. In addition,
and in agreement with Ref.~\cite{duyu+ni22}, the value of $g_0$
depends on density and structure. In effect, the calculation of
$\phi_{\mathrm{RCP}}$ in Ref.~\cite{zaccone22} is, unfortunately,
incorrect and an analytical calculation of $\phi_{\mathrm{RCP}}$ in
low dimensions remains an open problem. In hindsight, it may not be
surprising, that (i) $\phi_{\mathrm{RCP}}$ cannot be determined by an
infinite parameter like $g(\sigma)$ and (ii) local knowledge about
$g(\sigma)$ is not sufficient to define an isostatic structure but
instead the geometry in the immediate neighborhood, encoded in
$g_{\mathrm{div}}(r)$, must be specified.

\begin{acknowledgments}
  I should like to acknowledge funding by the \textsc{dfg} through
  KR\,4867/2 and interesting discussions with Olivier Coquand.
\end{acknowledgments}


\begin{thebibliography}{9}
\expandafter\ifx\csname natexlab\endcsname\relax\def\natexlab#1{#1}\fi
\expandafter\ifx\csname bibnamefont\endcsname\relax
  \def\bibnamefont#1{#1}\fi
\expandafter\ifx\csname bibfnamefont\endcsname\relax
  \def\bibfnamefont#1{#1}\fi
\expandafter\ifx\csname citenamefont\endcsname\relax
  \def\citenamefont#1{#1}\fi
\expandafter\ifx\csname url\endcsname\relax
  \def\url#1{\texttt{#1}}\fi
\expandafter\ifx\csname urlprefix\endcsname\relax\def\urlprefix{URL }\fi
\providecommand{\bibinfo}[2]{#2}
\providecommand{\eprint}[2][]{\url{#2}}

\bibitem[{\citenamefont{Zaccone}(2022{\natexlab{a}})}]{zaccone22}
\bibinfo{author}{\bibfnamefont{A.}~\bibnamefont{Zaccone}},
  \bibinfo{journal}{Phys. Rev. Lett.} \textbf{\bibinfo{volume}{128}},
  \bibinfo{pages}{028002} (\bibinfo{year}{2022}{\natexlab{a}}).

\bibitem[{\citenamefont{Chen and Ni}(2022)}]{duyu+ni22}
\bibinfo{author}{\bibfnamefont{D.}~\bibnamefont{Chen}} \bibnamefont{and}
  \bibinfo{author}{\bibfnamefont{R.}~\bibnamefont{Ni}},
  \emph{\bibinfo{title}{Comment on ``{E}xplicit {A}nalytical {S}olution for
  {R}andom {C}lose {P}acking in $d=2$ and $d=3$''}} (\bibinfo{year}{2022}),
  \eprint{arXiv:2201.06129}.

\bibitem[{\citenamefont{Charbonneau and Morse}(2022)}]{charbonneau+morse22}
\bibinfo{author}{\bibfnamefont{P.}~\bibnamefont{Charbonneau}} \bibnamefont{and}
  \bibinfo{author}{\bibfnamefont{P.~K.} \bibnamefont{Morse}},
  \emph{\bibinfo{title}{Comment on ``{E}xplicit {A}nalytical {S}olution for
  {R}andom {C}lose {P}acking in $d = 2$ and $d = 3$''}} (\bibinfo{year}{2022}),
  \eprint{arXiv:2201.07629}.

\bibitem[{\citenamefont{Blumenfeld}(2022)}]{blumenfeld22}
\bibinfo{author}{\bibfnamefont{R.}~\bibnamefont{Blumenfeld}},
  \emph{\bibinfo{title}{Comment on ``{E}xplicit {A}nalytical {S}olution for
  {R}andom {C}lose {P}acking in $d=2$ and $d=3$''}} (\bibinfo{year}{2022}),
  \eprint{arXiv:2201.10550}.

\bibitem[{\citenamefont{Zaccone}(2022{\natexlab{b}})}]{zaccone22e}
\bibinfo{author}{\bibfnamefont{A.}~\bibnamefont{Zaccone}},
  \bibinfo{journal}{Phys. Rev. Lett.} \textbf{\bibinfo{volume}{129}},
  \bibinfo{pages}{039901} (\bibinfo{year}{2022}{\natexlab{b}}).

\bibitem[{\citenamefont{Charbonneau et~al.}(2017)\citenamefont{Charbonneau,
  Kurchan, Parisi, Urbani, and Zamponi}}]{charbonneau+kurchan17}
\bibinfo{author}{\bibfnamefont{P.}~\bibnamefont{Charbonneau}},
  \bibinfo{author}{\bibfnamefont{J.}~\bibnamefont{Kurchan}},
  \bibinfo{author}{\bibfnamefont{G.}~\bibnamefont{Parisi}},
  \bibinfo{author}{\bibfnamefont{P.}~\bibnamefont{Urbani}}, \bibnamefont{and}
  \bibinfo{author}{\bibfnamefont{F.}~\bibnamefont{Zamponi}},
  \bibinfo{journal}{Annu. Rev. Condens. Matter Phys.}
  \textbf{\bibinfo{volume}{8}}, \bibinfo{pages}{265} (\bibinfo{year}{2017}).

\bibitem[{\citenamefont{Kanwal}(1998)}]{kanwal98}
\bibinfo{author}{\bibfnamefont{R.~P.} \bibnamefont{Kanwal}},
  \emph{\bibinfo{title}{Generalized {F}unctions: {T}heory and {T}echnique}}
  (\bibinfo{publisher}{Birkh{\"a}user}, \bibinfo{address}{Boston},
  \bibinfo{year}{1998}), chap. \bibinfo{chapter}{3.3}.

\bibitem[{\citenamefont{Torquato}(2018)}]{torquato18}
\bibinfo{author}{\bibfnamefont{S.}~\bibnamefont{Torquato}},
  \bibinfo{journal}{J. Chem. Phys.} \textbf{\bibinfo{volume}{149}},
  \bibinfo{pages}{020901} (\bibinfo{year}{2018}).

\bibitem[{\citenamefont{Zaccone}(2022{\natexlab{c}})}]{zaccone22r}
\bibinfo{author}{\bibfnamefont{A.}~\bibnamefont{Zaccone}},
  \emph{\bibinfo{title}{Reply to {W}. {T}. {K}ranz on ``{E}xplicit {A}nalytical
  {S}olution for {R}andom {C}lose {P}acking in d=2 and d=3''}}
  (\bibinfo{year}{2022}{\natexlab{c}}), \eprint{arXiv:2205.00720}.

\end{thebibliography}

\end{document}